# Ecosystems, from life, to the Earth, to the Galaxy

Michael G Burton

School of Physics, University of New South Wales, Sydney, NSW 2052, Australia

and

School of Cosmic Physics, Dublin Institute for Advanced Studies, 5 Merrion Square, Dublin 2, Ireland

Telephone:    +61-2-9385-5618

Fax:          +61-2-9385-6060

Email:        M.Burton@unsw.edu.au

*Abstract*

Ecosystems are systems where energy flows and material cycles are maintained in an apparently stable, but non-equilibrium state through a process of self-regulation. Such a definition does just apply to biological systems, it can also apply to systems that involve entirely physical processes. We discuss how four systems, each operating on very different spatial and temporal scales, each exhibit these features of an ecosystem. These are, in order of increasing magnitude, the cell, the forest, the Earth and the Galaxy. In particular, we discuss how the process of star formation across the spiral arms of galaxies works as an ecosystem. The carbon abundance plays a crucial role in both the self-regulation and the evolution of the system. We suggest that spiral galaxies may be the first ecosystems to form in the Universe after the Big Bang.

## Introduction

Ecosystems are usually associated with life on the Earth, a community of organisms interacting and evolving in a manner determined by the totality of their biological and physical environment. Driven by an energy source, typically the Sun, material is re-cycled and the system is in a self-regulating, non-equilibrium state. Ecosystems are not just confined to geographical regions of the Earth, however. The community of organisms that make up the cells of living creatures acts as an ecosystem. The Earth itself can be regarded as an ecosystem. The Gaia hypothesis (Lovelock and Margulis 1974) argues that evolution through natural selection, from the time of the earliest bacteria, and the role of organisms in the chemical and physical cycles of the biosphere, have played a significant role in the self-regulation of both the contents and the physical state of the land, oceans and atmosphere. On a larger scale, our Milky Way galaxy can also be regarded as an ecosystem. The exchanges of matter and energy between the stars and the interstellar medium over the life cycle of stars, drive the process of star formation across the spiral arms of the Galaxy in an apparently stable, but non-equilibrium state. It also produces the elements, and possibly the organic molecules, that are a pre-cursor to life.

This paper discusses how the concept of an ecosystem can be used to provide insight into the workings all these structures, operating at vastly disparate scales. The themes of energy flows and material cycles, in systems operating far from thermal equilibrium, yet able to self-regulate in an autonomous manner, are common to them all. They are what defines an ecosystem. In particular, we aim to show how the processes that operate in a spiral galaxy like our own Milky Way fulfil these criteria. Indeed, despite the tremendous spatial and time scales they operate on, at least compared to those we are familiar with on the Earth, the spiral galaxies may be regarded as the simplest examples of ecosystems in our Universe, operating entirely by physical mechanisms without any biological input. They are also the first ecosystems to evolve and establish themselves after the formation of the Universe.

## The Forest

The community of organisms that live, and depend on each other for their existence, in a forest is perhaps the most familiar example of an ecosystem (e.g. see the entry on ecosystems in the Encyclopaedia Britannica). We typically associate this with the plants and animals of the forest, but it also includes the micro-organisms found within and around them, as well as the environment which they inhabit. These ecosystems contain both the physical components that define their environment, as well as the biological components that form the community of organisms living in it. This community forms a food web comprising of producers, consumers and decomposers. The producers are mainly the green plants which take their energy directly from sunlight, using it to combine $CO_2$ and $H_2O$ directly to form carbohydrates and $O_2$ in the process of photosynthesis. The consumers (the animals) ingest these organisms, or other consumers. Through respiration, breathing in $O_2$, they derive energy from the break up of carbon bonds when the $O_2$ reacts with the carbohydrates, returning $CO_2$ to the air. The decomposers (micro-organisms) break down the waste organic products of the producers and consumers (including when they die), releasing the raw materials for recycling. Energy has flowed through the system, having been supplied from outside the system, by the Sun. Nutrients have cycled around it through this food web. Biodiversity enhances this process and provides stability, e.g. in case one particular species dies out. The processes operating are self-regulating and maintain a balance between the numbers of particular organisms. However, while over a few years the system may appear quasi-stable (aside from seasonal cycles and yearly climatic variations), both the type and number of organisms do vary with the time. The ecosystem evolves, by a process of natural selection in the biological community, in a completely autonomous manner.

### *The Cell*

The cell is the fundamental unit of biological organisation, and contains all the biochemical machineries needed for metabolism and reproduction. However the cell is also a community of organisms, its constituent parts forming an ecosystem whose interactions are essential for their continued existence (e.g. Thomas 1974). Eukaryotic cells are surrounded by a membrane. Through this they absorb nutrients, and expel waste and the products they manufacture, such as proteins and fats. The interior of the cell is filled with a fluid medium, the cytoplasm, in which a variety of organelles co-exist. These are the cell's living machinery and form the community of organisms which act together in the ecosystem. They include the nucleus carrying the genetic information, the mitochondria for energy production, the chloroplasts (in plants) for photosynthesis, lysosomes for the cell's defence, ribosomes for manufacturing proteins, the endoplasmic reticulum for their transportation within the cell and the golgi apparatus for packaging them for distribution outside it. Parts of the cell, such as the mitochondria, even have their own independent DNA, passed along entirely through the egg cells of the host organism's maternal line, and replicating on their own, independent of replication of the cell. This community of organisms acts autonomously and continuously, converting sunlight into chemical bonds and releasing energy. They process nutrients such as salts and proteins, and cycle them through their environment. The cell has evolved to its present forms, and its parts have come together through the process of endosymbiosis (e.g. Margulis 1970).

### *The Earth*

The Earth has been compared to a single cell (Thomas 1974) and described as "symbiosis as seen from space" (Margulis 1998). It is a coupled system consisting of the land, the oceans, the atmosphere, the climate and the biota. No part can act completely independently of another. For instance, even geological processes are influenced by carbon deposition in limestone, built-up ultimately from that provided by the biota, as part of the endogenic cycle. The Gaia hypothesis (Lovelock 1995) has often been misinterpreted as implying that the Earth is alive or is self-directed, but in fact it says that it is a self-regulating system, operating autonomously through a series of interacting bio-geo-chemical cycles. The maintenance of conditions comfortable for life over nearly 4 billion years, despite the solar energy output having increased by 25% in that time, attests to this. The evolution of the biota has been largely responsible for the change in the atmospheric composition at the end of the Archean (first appearance of $O_2$) and the Proterozoic eras (build up of $O_2$ to current levels) of geological history, roughly 2.5 and 0.6 billion years ago, respectively. The energy which drives the system flows, ultimately, from the Sun (though geothermal energy, arising from radioactivity, can play a role in localised regions). The chemicals of life all pass through major biogeochemical cycles, spending most of their time in reservoirs and passing through exchange pools (e.g. carbon, as found in limestone rocks and when hosted in biological organisms). The system is in a non-equilibrium state, as evident by the presence of the combustible molecules methane and oxygen in the atmosphere. These react together and therefore must continually be replaced to maintain equilibrium. This therefore implies a means of self-regulation to keep the levels constant. The system is complex and its constituents are continually in competition. Yet the great diversity it also exhibits serves to provide stability, and a means to respond to cataclysmic events (as evident through several mass extinctions in the geological record, and the subsequent recovery and diversification of species). The Earth has evolved, and is continuing to evolve, through the process of natural selection. As with the cell and the forest, these are the traits that an ecosystem exhibits.

## *The Galaxy*

Star Formation

Our Milky Way galaxy displays the same characteristics that we associate with ecosystems if we view it on appropriately large scales of distance and time. This can most clearly be seen if we take an imaginary trip outside our Galaxy, to view it from afar. Seen face-on the Galaxy appears as a spiral, roughly 100,000 light years in extent. This spiral shape actually serves to hinder our appreciation of the true structure, which is that of a thin disk (a few hundred light years thick) full of gas and stars, enveloped in a roughly spherical halo, sparsely populated with stars. The spiral shape which attracts our attention, however, is where the galactic ecosystem is at work. Here the process of star formation is active, a process that drives the energy flows and material cycles that are necessary to set up an ecosystem. Viewed on timescales of millions of years the galaxy would appear 'alive' in much the same way as the Earth would if viewed at similar intervals so that we could see continental drift at work. The spiral arms are a 'compression wave', where the gas density is increased. They are almost stationary with respect to the rotation of the stars and gas around the galactic centre. Gas thus flows into the spiral arms, is compressed and a portion emerges as new stars (the stars themselves flow unimpeded through the spiral arms). The number of stars which form is critically dependent on the local environment. Feedback processes control how efficiently and at what rate this can occur. Yet it is a process which must have been operating over most of the 10 billion year history of our Galaxy to build up the stellar population that resides in it today.

Stars form from the collapse of clouds of molecular gas under the influence of gravity. The clouds are prevented from collapsing by a number of internal support mechanisms, such as pressure, rotation, magnetic fields and turbulence. They require a triggering mechanism to overcome this, which may be the compression wave of an approaching spiral arm. Once star formation starts in a cloud it is then dominated by the most massive stars which form within it. These stars, with masses of between 10 and 100 times that of our own Sun, shine with a luminosity which is a thousand to a million times greater than it. They create ionized bubbles around them ('HII regions'), which rapidly expand to envelope and disrupt the surrounding molecular cloud. The spiral arms are thus lit up by the massive stars forming in them. Yet they only make up a tiny percentage of the number of stars that are forming at any given time, or indeed of the total number of stars in the Galaxy. The rate of star formation in a particular cloud, and thus the efficiency of star formation, depends on the local environment of the cloud, and on the stars already forming in it. Massive stars not only emit prodigious amounts of ionizing radiation, but they also eject tremendous winds, streams of atoms moving at speeds of hundreds of kilometres per second. These drive shock waves through the surrounding molecular clouds. Such interactions can serve to cause further star formation, compressing the gas and triggering collapse. Or they can hinder it, breaking up the molecular cloud and providing turbulent support against further collapse, thus preventing any more stars from forming.

Carbon Catalysis

The rate of star formation, and the mass of the stars that form, is also strongly influenced by the chemical composition of the clouds. To collapse efficiently a cloud must be able to release the gravitational energy it loses quickly, otherwise it heats up, generating thermal pressure to support the cloud against further collapse. It releases this energy by emitting radiation through the lines of the atoms and molecules present in it, cooling the cloud. How efficiently a cloud cools is determined by the elemental abundance of the cloud, particularly the presence of carbon. This is because the dominant element, hydrogen, does not provide an efficient line cooling mechanism when the temperature drops below ~1000°C. Carbon, either in the singly ionized form emitting at a wavelength of 158μm, or in molecular form emitting several lines in the millimetre portion of the spectrum, provides the dominant cooling mechanism across a wide range of density conditions prevalent in molecular clouds (e.g. Tielens & Hollenbach 1985). The abundance of carbon in the Galaxy has been increasing as the Galaxy has aged. Thus its influence in regulating the rate of star formation, the resultant distribution of stellar masses, and even the processes taking place during the lives of individual stars, has changed over the history of the Galaxy.

Carbon, and all heavier elements in the Universe (up to the element iron), were created by the process of nucleosynthesis in stars. Initially, some 13 billion years ago and following the Big Bang, the Universe consisted of roughly 76% hydrogen and 24% helium and a trace of lithium. The elements that are essential to life were later created in stars, and then released into the interstellar medium, much of it through the supernova explosions that only massive stars go through as they die. The carbon that is found in the interstellar medium, though, while produced by nucleosynthesis in stars, is probably mostly placed into the interstellar medium by powerful winds, driven by radiation pressure from the massive stars (e.g. Gustafsson et al. 1999), having been dredged up from their cores, rather than being released in a supernova explosion at the end of that star's life. It is literally true to say that "we are stardust", for the elements of our bodies have passed through several generations of stars before they accumulated in the molecular cloud that eventually formed our Solar

System, our Earth, and us. The massive stars have created the raw material for life. It is the debris from the star-gas cycle of the Galaxy.

## The Galactic Ecosystem

The stars themselves are far from thermal equilibrium, with the energy flows from their surface's radiating into cold space. The stars are stable, but non-equilibrium systems, that have their own cycle of birth, life and death. The timescale for the Galactic ecology is determined by the rate of star formation and the lifetime of the most massive stars (a few million years). This ecology must have existed, though in gradually changing form, over the life of the Galaxy. It is driven by the energy flows from the massive stars, and the material cycle through these same stars. Carbon, and heavier elements, are created in the massive stars, and released through winds and supernova explosions. They cycle between the various phases of the interstellar medium, before again being incorporated into stars and, in some cases, planetary systems and life. Further star formation in a molecular cloud is self-regulated by the massive stars already forming, and by the cooling agents which are already present in it. These agents gradually change as the elemental abundances, particularly of carbon, increase as the Galaxy evolves. A feedback process operates which both regulates the star formation rate and the stellar mass spectrum (ie. the range of stellar types and the number of stars within each type).

The elements of an ecosystem are therefore in operation in our Galaxy; energy flows and materials cycles, driven by the massive stars, in systems far from thermal equilibrium, and whose end result is determined by a process of self-regulation. Despite this delicate balance and interplay between many aspects of the star formation cycle, the system has continued to operate for ten billion years. It remains to be asked whether it is evolving with time, the final characteristic of an ecosystem, and a necessity to maintain a system operating far from equilibrium? Or whether it is simply an autocatalytic cycle, always returning to the same starting point? Here the comparison with ecosystems on Earth is weaker, for evolution in the Galaxy is limited compared to the tremendous biodiversity it has produced on Earth. Star formation produces a range of stellar masses, with the initial mass function (IMF) determining how many stars of a particular mass are found. The different masses can be regarded as analogous to the different species within an ecosystem. The life story of an individual star is determined largely by its mass, and is very different for low mass (ie. Solar mass and below) and high mass (> 10 Solar mass) stars. While astronomers often refer to this as stellar evolution, this is an inappropriate use of the word, as it is used to describe the life processes of individual stars, not their changes from one generation to the next. The luminosity, the life-span, the type of nuclear reactions at work within the core, and the manner of energy transport through the star, are all strongly dependent on the stellar mass. They also depend on secondary characteristics such as the initial chemical composition, a quantity which changes from generation to generation as the Galaxy ages. The amount of mass lost by the most massive stars (those greater than 25 times the Solar mass), having been enriched through nucleosynthesis before being returned to the interstellar medium, is strongly dependent on the initial carbon abundance of the star (Maeder 1992).

## The Evolution of the Initial Mass Function (IMF)

The average value of the IMF can be measured. It takes the form $dN/dM \propto M^{-2.3}$ for most stars in our Galactic neighbourhood (where $N(M)$ is the number of stars of mass M, as initially deduced by Salpeter 1955, but see Zinnecker, McCaughrean & Wilking 1993 for a recent discussion), and is heavily weighted towards low mass stars. Why the IMF takes this form is not understood, although explanations have been proposed (e.g. Silk 1995). It is connected with the manner in which stars accrete material, and their subsequent interaction with outflows that are generated by the need to shed angular momentum. However, the abundance of the elements within a cloud, particularly that of carbon, certainly affects the IMF due to its role in the cooling of the cloud as it collapses. At early times in our Galaxy the IMF would likely have been weighted towards higher mass stars than it is today, impeding the collapse at low masses (e.g. Larson 1998). This is because the lower elemental abundances result in the molecular clouds then being, on average, hotter than today (due to more inefficient cooling processes operating, such as through hydrogen molecules). Higher masses are thus needed to overcome the thermal pressures in order to collapse. As the Galaxy ages, and the carbon abundance increases, the resulting IMF can be expected to change, though in ways that are not yet properly understood.

Thus the Galaxy is displaying the first symptoms of what we recognise as evolution, through the change in the IMF with age, and therefore in the community of stars that exist within it. New types of stars, appear as the Galaxy evolves. This occurs due to a limited form of information passage from one generation of stars to the next, through the increase in their initial elemental abundances (astronomers refer to this as an increase in "metallicity", though this is a singularly inappropriate use of the word, as it is applied to all elements heavier than lithium!). The changing carbon abundance in star forming clouds then affects the subsequent evolution of the Galaxy. The stars that form in one generation influence the type and number of stars that form in the next, through the feedback mechanisms that operate and the resultant increase in elemental abundance of the interstellar medium. Thus, the system shows a limited form of natural selection in its evolution. We have an analogy with the concept of evolution by descent which we associate with biological systems, though in a greatly simplified manner. There is self-replication (ie. more stars are created) with variation (ie. a changing IMF and stellar behavioural properties, as the elemental abundances build up in successive generations), and these continue to replicate with variation, at least until all the raw material has been exhausted. This is the basic process we also associate with biological evolution (e.g. Ehrlich 1991), though there are, of course, significant differences between the evolutionary processes at work in each system.

The diversity exhibited by the new generations of stars is vastly simplified compared to the many forms that biological evolution takes. The information passed from one generation to the next, through the elemental abundances, is extremely limited compared to the richness that is transferred through DNA. There is also no real parallel to the concept of extinction, unless it is with regard to massive stars with low elemental abundances, which will cease to exist as a galaxy ages. Lower mass stars, with a wide range of elemental abundances, will all co-exist in the Galaxy (having lifespans greater than the age of the Galaxy), the new generation of stars simply joining the older generation, not replacing them. For higher mass stars, however, the low elemental abundances in the first generation will be replaced by stars of ever higher abundances (as their lifespan is very much less than the age of the Galaxy). There is also a limited analogy with competition between species as seen in a biological system, though there is certainly competition between the protostars within a molecular cloud with regard to the supply of raw materials they need to form. This occurs through the gravitational interactions between the stars, determining which star, if any, a particular atom or molecule will be incorporated into.

The Galaxy thus operates as an ecosystem, but involves entirely physical mechanisms and not biological ones. Energy flows and material cycles, self-regulated but in a non-equilibrium state, are occurring. There is also a limited form of evolution, as the community of stars within the Galaxy changes with time. Spiral galaxies are the first ecosystems to have formed in the Universe, and the simplest such examples.

Examining the Case for Carbon Self-Regulation

For this description of the Galactic ecosystem to have true merit it needs to help us understand particular phenomena or help to predict their behaviour in certain circumstances. The suggestion that star formation has been occurring in a self-regulating manner over the ten billion years of galactic history, is most open to criticism. We have hypothesised that it is the massive stars, and the role of the heavy elements, particularly carbon, that is central to this process. This can be subject to scrutiny, as the origin of the elements, and their distribution through the Galaxy, is not yet well understood. Currently the evidence does indeed favour the production of carbon through winds from massive (> 10 Solar masses) stars as being the dominant contributor (e.g. see Henry, Edmunds & Koppen 2000). This provides an effectively instantaneous production of carbon compared to galactic timescales, as it does not even require the star to complete its life before enriching the gas. Thus, as the carbon abundance increases, then so will the regulatory mechanism alter. This can therefore can be studied to see if the ecosystem we have described could be maintained over time. For instance, it has not always been accepted that carbon production is dominated by the massive star winds. Carbon can be produced, via the triple-alpha reaction, by all stars greater in mass than Solar. The lower mass stars can supply carbon into the interstellar medium via mass loss during the 'planetary nebula' portion of their 'red giant' phase of their lives (which occurs near the end of stellar life, when nuclear helium-burning is taking place in the core and the outer envelope shed). Since low mass stars greatly outnumber high mass ones, this could provide the dominant method of carbon-enrichment. If low mass stars do dominate, then the means of self-regulation of the system, which is driven by the massive stars, would fail and the ecosystem, as described here, would not operate. There have been shifting views with regards to the relative roles of high and low mass stars in this enrichment process (see Gustafsson et al. 1999), although high-mass stars are currently favoured. It does seem clear, however, that mass rate losses are very sensitive to the carbon abundance in the star (being greater at higher abundance, due to the increased opacity in the surface layers of those stars, improving the efficiency of the radiative driving mechanism for their winds; see Maeder 1992). On the other hand, as the Galaxy ages, and the number of (long-lived) low mass stars increases, the relative importance of the lower mass stars will likely increase (see Liang, Zhao & Shi 2001), possibly removing the self-regulatory mechanism of carbon production from the massive stars. This is an active area of research, and if the Galactic ecosystem hypothesis is to be supported, then carbon enrichment from the massive stars, those driving the star formation cycle across the spiral arms, would have to be found to be dominant mechanism while the ecosystem is operational.

## *Conclusions*

Ecosystems are systems operating far from thermal equilibrium, driven by flows of energy from one part of the system to another, whose raw materials are continually re-cycled as they pass through the components of the system. Ecosystems operate autonomously, by a process of self-regulation. Their flows of energy mean they cannot remain static with time. Although autocatalytic cycles may exist at times (when the system returns to its starting state), as the Universe evolves so must the system if it is to be maintained. That evolution occurs by a process of natural selection, though the use of that term should be widened to include purely physical mechanisms as well as those involving biological organisms. These common traits are exhibited, though in vastly different ways, by the four systems we have discussed; the cell, the forest, the Earth and the Galaxy. These systems each operate on very different spatial and temporal scales from each other. Each system also shows structure over a range of scales, they are 'critical systems'. A system showing structure on all scales cannot be in thermodynamic equilibrium, but must continually change with time, as energy flows through it, from one scale to another. This provides a link between the ecosystems we have described, as they cannot act completely independently of each other. The Universe is displaying the characteristics of self-organisation through the ecosystems operating at different levels within it (see Smolin 1997 for a fuller discussion of this behavioural aspect). It has produced the Galactic ecosystem, which has produced the Earth, which has produced life.

Our Galaxy, and spiral galaxies in general, must be the first ecosystems to form within the Universe, within the first 3 billion years of its existence. They are thus the simplest such systems to exist within it, with their operation determined entirely by physical processes. Yet there is still much we don't understand about them, for instance the cause of the initial mass function which determines the range and types of stars that exist. Carbon plays a central role in both the self-regulation and the evolution of the system. By studying such an ecosystem, and the essential behavioural elements within it, we may hope to gain insight into the vastly more complex ecosystems that involve life.